\documentclass[10pt,letter,prl,twocolumn,nofootinbib,hypertext]{revtex4-1}
\pdfoutput=1
\usepackage{ifthen}
\newboolean{uprightparticles}
\setboolean{uprightparticles}{false} %Set to true to get roman particle symbols

\usepackage{amssymb}
\usepackage{amsfonts}
\usepackage{upgreek}
\usepackage{xspace}
\usepackage{color}
\usepackage{amsmath}
\usepackage{graphicx}
\usepackage{bm}
\usepackage{ctable}
\usepackage{placeins}
\usepackage{multirow}
\usepackage{dcolumn}
\graphicspath{{./figs/}} % Make Latex search fig subdir for figures
\usepackage{hyperref}
\hypersetup{%
 colorlinks=false, linktocpage=false, pdfborder={0 0 0}, pdfstartpage=1, 
 pdfstartview=FitV,% 
  }

\newboolean{articletitles}
\setboolean{articletitles}{true} 

\newboolean{pdflatex}
\setboolean{pdflatex}{true} % use this if using non-eps figures

\newcommand{\BRof}[1]{\ensuremath{{\cal B}(#1)}\xspace}
\newcommand{\Bsmumu}{\ensuremath{B^0_s \to\mu^+\mu^-}\xspace}
\newcommand{\Bdmumu}{\ensuremath{B^0\to\mu^+\mu^-}\xspace}
\newcommand{\Bmm}{\ensuremath{B^{0}_{(s)}\to\mu^+\mu^-}\xspace}

\newcommand{\BspiK}{\ensuremath{B^0_s\to\pi^+K^-}\xspace}

\newcommand{\Bdpipi}{\ensuremath{\Bd\to\pi^+\pi^-}\xspace}
\newcommand{\bpimumu}{\ensuremath{B^{0(+)} \to \pi^{0(+)} \mu^+ \mu^-}\xspace}
\newcommand{\BdPiMuNu}{\ensuremath{\ensuremath{B^0}\to \pi^- \mu^+ \nu_\mu}\xspace}

\def\B       {\ensuremath{B}\xspace}
\def\Bd      {\ensuremath{B^0}\xspace}
\def\Bs      {\ensuremath{B^0_s}\xspace}
\newcommand{\mBd}{\ensuremath{m_{\Bd}}\xspace}
\newcommand{\mBs}{\ensuremath{m_{\Bs}}\xspace}

\def\lhcb {LHCb\xspace}
\newcommand{\Bhh}{\ensuremath{B^0_{(s)}\to h^+{h}^{\prime -}}\xspace}

\newcommand{\Bmumu}{\ensuremath{B^0_{(s)}\to \mu^+\mu^-}\xspace}
\newcommand{\BuJpsiK}{\ensuremath{B^+\to J/\psi K^+}\xspace}
\newcommand{\BsJpsiPhi}{\ensuremath{B^0_s\to J/\psi \phi}\xspace}
\newcommand{\BdKpi}{\ensuremath{B^0\to K^+\pi^-}\xspace}
\newcommand{\BsKK}{\ensuremath{B^0_s\to K^+K^-}\xspace}
\newcommand{\bbdim}{\ensuremath{b\bar{b}\to \mu^+ \mu^- X}\xspace}
\newcommand{\CLsb}{\ensuremath{\textrm{CL}_{\textrm{s+b}}}\xspace}
\newcommand{\CLs}{\ensuremath{\textrm{CL}_{\textrm{s}}}\xspace}
\newcommand{\CLb}{\ensuremath{\textrm{CL}_{\textrm{b}}}\xspace}

\newcommand{\gevc}{\ensuremath{{\mathrm{\,Ge\kern -0.1em V\!/}c}}\xspace}
\newcommand{\mevc}{\ensuremath{{\mathrm{\,Me\kern -0.1em V\!/}c}}\xspace}
\newcommand{\gevcc}{\ensuremath{{\mathrm{\,Ge\kern -0.1em V\!/}c^2}}\xspace}
\newcommand{\gevgevcccc}{\ensuremath{{\mathrm{\,Ge\kern -0.1em V^2\!/}c^4}}\xspace}
\newcommand{\mevcc}{\ensuremath{{\mathrm{\,Me\kern -0.1em V\!/}c^2}}\xspace}

\newcommand\TTstrut{\rule{0pt}{3.2ex}}
\newcommand\Bstrut{\rule[-1.2ex]{0pt}{0pt}}
\newcommand\BBstrut{\rule[-1.8ex]{0pt}{0pt}}

\def\invfb   {\ensuremath{\mbox{\,fb}^{-1}}\xspace}
\newcommand{\tev}{\ensuremath{\mathrm{\,Te\kern -0.1em V}}\xspace}
\newcommand{\CL}{CL\xspace}
\usepackage{hyperref}
\usepackage[all]{hypcap}
\usepackage{mciteplus}

\begin{document}
%------------------

\begin{titlepage}
\pagenumbering{roman}

% Header ---------------------------------------------------
\vspace*{-1.5cm}
\centerline{\large EUROPEAN ORGANIZATION FOR NUCLEAR RESEARCH (CERN)}
\vspace*{1.0cm}
\hspace*{-0.5cm}
\begin{tabular*}{\linewidth}{lc@{\extracolsep{\fill}}r}
\ifthenelse{\boolean{pdflatex}}% Logo format choice
{\vspace*{-2.7cm}\mbox{\!\!\!\includegraphics[width=.14\textwidth]{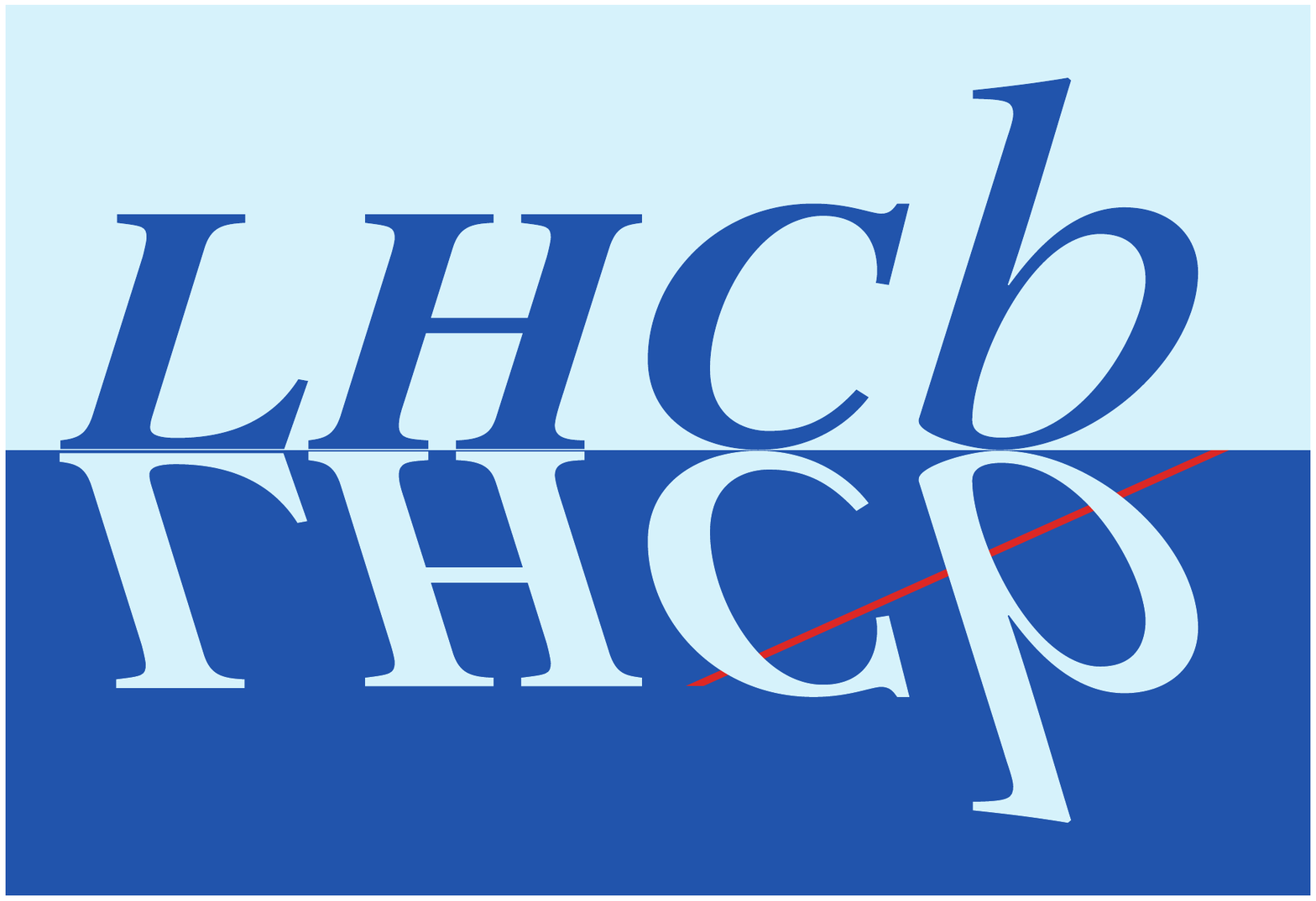}} & &}%
{\vspace*{-1.2cm}\mbox{\!\!\!\includegraphics[width=.12\textwidth]{lhcb-logo.eps}} & &}%
\\
 & & CERN-PH-EP-2012-335 \\  
 & & LHCb-PAPER-2012-043 \\  
 & & November 12, 2012 \\ 
 & & \\
% not in paper \hline
\end{tabular*}

\vspace*{2.0cm}

% Title --------------------------------------------------
{\bf\boldmath\huge
\begin{center}
 First evidence for the decay \mbox{\boldmath $B^0_s \rightarrow \mu^+\mu^-$}
\end{center}
}

\vspace*{1.0cm}

% Authors -------------------------------------------------
\begin{center}
The LHCb collaboration
%\footnote{Authors are listed on the following pages.}
\end{center}

%\vspace{\fill}
\vspace*{5mm}
%\input{LHCb_authorlist_lhcb}

%%%%%%%%%%%%%%%%%%%%%%%%%%%%%%%%%%%%%%%%%%
%\centerline{\large\bf LHCb collaboration}
\begin{flushleft}
\small
R.~Aaij$^{38}$, 
C.~Abellan~Beteta$^{33,n}$, 
A.~Adametz$^{11}$, 
B.~Adeva$^{34}$, 
M.~Adinolfi$^{43}$, 
C.~Adrover$^{6}$, 
A.~Affolder$^{49}$, 
Z.~Ajaltouni$^{5}$, 
J.~Albrecht$^{35}$, 
F.~Alessio$^{35}$
% ...

{\footnotesize \it
% ...
$ ^{5}$Clermont Universit\'{e}, Universit\'{e} Blaise Pascal, CNRS/IN2P3, LPC, Clermont-Ferrand, France\\
$ ^{6}$CPPM, Aix-Marseille Universit\'{e}, CNRS/IN2P3, Marseille, France\\
% ...
$ ^{11}$Physikalisches Institut, Ruprecht-Karls-Universit\"{a}t Heidelberg, Heidelberg, Germany\\
% ...
$ ^{33}$Universitat de Barcelona, Barcelona, Spain\\
$ ^{34}$Universidad de Santiago de Compostela, Santiago de Compostela, Spain\\
$ ^{35}$European Organization for Nuclear Research (CERN), Geneva, Switzerland\\
% ...
$ ^{38}$Nikhef National Institute for Subatomic Physics, Amsterdam, The Netherlands\\
% ...
$ ^{43}$H.H. Wills Physics Laboratory, University of Bristol, Bristol, United Kingdom\\
% ...
$ ^{49}$Oliver Lodge Laboratory, University of Liverpool, Liverpool, United Kingdom\\
% ...
$ ^{n}$LIFAELS, La Salle, Universitat Ramon Llull, Barcelona, Spain\\
% ...
}
\end{flushleft}
%%%%%%%%%%%%%%%%%%%%%%%%%%%%%%%%%%%%%%%%%%

\cleardoublepage

% Abstract -----------------------------------------------
\begin{abstract}
\noindent 
A search for the rare decays \Bsmumu and \Bdmumu is performed  using data collected in 2011 and  
2012 with the LHCb experiment at the Large Hadron Collider.
The data samples comprise 1.1\invfb of proton-proton collisions at $\sqrt{s} = 8$ TeV
and 1.0\invfb at $\sqrt{s}=7$ TeV. 
We observe an excess of \Bsmumu candidates with respect to the background expectation.
The probability that the background could produce such an excess or larger is $5.3\times 10^{-4}$
corresponding to a signal significance of 3.5 standard deviations.
A maximum-likelihood fit gives a branching fraction of \BRof \Bsmumu = $(3.2^{\,+1.5}_{\,-1.2}) \times 10^{-9}$,
where the statistical uncertainty is  95\,\% of the total uncertainty.
This result is in agreement with the Standard Model expectation.
The observed number of $B^0 \rightarrow \mu^+ \mu^-$ candidates
is consistent with the background expectation, giving an upper limit of
\BRof \Bdmumu $< 9.4 \times 10^{-10}$ at 95\,\% confidence level.

\bigskip
\begin{center}
{Submitted to Physical Review Letters}
\end{center}

\end{abstract}

\maketitle
\end{titlepage}

\pagestyle{plain}
\setcounter{page}{1}
\pagenumbering{arabic}

%----------------
% Introduction
%----------------
\noindent
The rare decays \Bsmumu and \Bdmumu 
are highly suppressed in the Standard Model (SM).
Precise predictions of their branching fractions,
\BRof \Bsmumu = $(3.23 \pm 0.27) \times 10^{-9}$ and \BRof \Bdmumu = $(1.07 \pm 0.10) 
\times 10^{-10}$~\cite{Buras2012} 
make these modes powerful probes in the search for deviations from the
SM, especially in models with a non-standard Higgs sector.
Taking the measured finite width difference of the \Bs system~\cite{LHCb-CONF-2012-002} into 
account~\cite{deBruyn:2012wk},
the time integrated branching fraction of \Bsmumu that should be compared to the experimental value 
is $(3.54 \pm 0.30) \times 10^{-9}$.

Previous searches~\cite{d0_PLB,cdf_prl,cms2,atlas,lhcbpaper3}
already constrain possible deviations from the SM predictions.
The lowest published limits are 
\mbox{\BRof \Bsmumu $< 4.5 \times 10^{-9}$} and \mbox{\BRof \Bdmumu $< 1.0 \times 10^{-9}$} 
at 95\,\% confidence level (\CL) from the LHCb collaboration using 1.0\invfb of data 
collected in $pp$ collisions in 2011 at $\sqrt{s}=7$ TeV~\cite{lhcbpaper3}.
This Letter reports an update of this search with 1.1 fb$^{-1}$ of data
recorded in 2012 at $\sqrt{s} = 8$ TeV.

%--------------------------
%-- analysis structure
%--------------------------
The analysis of 2012 data is similar to that described in Ref.~\cite{lhcbpaper3} with two main improvements: 
the use of particle identification to select  \Bhh (with $h^{(\prime)} = K, \pi$) 
decays used to calibrate the geometrical and kinematic variables, and
a refined estimate of the exclusive backgrounds. 
To avoid potential bias, the events in the signal region were not examined until all the
analysis choices were finalized. The updated estimate of the exclusive backgrounds is also applied 
to the 2011 data~\cite{lhcbpaper3} and the results re-evaluated.
The results obtained with the combined 2011 and 2012 datasets supersede those of Ref.~\cite{lhcbpaper3}.

%--------------------------
%-- detector and trigger
%--------------------------
The \lhcb detector is a single-arm forward
spectrometer covering the pseudorapidity range \mbox{$2<\eta<5$}, and is 
described in detail in Ref.~\cite{LHCbdetector}.  
The simulated events used in this analysis are produced using the software described 
in Refs.~\cite{Sjostrand:2006za, Lange:2001uf, Allison:2006ve, Agostinelli:2002hh, 
Golonka:2005pn, LHCb-PROC-2011-005, LHCb-PROC-2011-006}. 

Candidate \Bmumu events are required to be selected by a hardware and a
subsequent software trigger. The candidates are predominantly selected by single and
dimuon triggers~\cite{LHCb-PUB-2011-017} and, to a smaller extent,
by a generic $b$-hadron trigger~\cite{LHCb-PUB-2011-016}.
Candidate events in the \BuJpsiK control channel, with $J/\psi \to \mu^+ \mu^-$
(inclusion of charged conjugated processes is implied throughout this Letter), 
are selected in a very similar way, the only difference
being a different dimuon mass requirement in the final software trigger. 
The \Bhh decays are predominantly selected by a hardware trigger
based on the calorimeter transverse energy and subsequently by a generic $b$-hadron software trigger.

%---------------
%-- selection
%---------------
The \Bmumu candidates are selected by requiring 
two high quality muon candidates~\cite{muonid} 
displaced with respect to any $pp$ interaction vertex (primary vertex, PV), 
and forming a secondary vertex (SV) with a $\chi^2$ per degree of freedom smaller than 9
and separated from the PV in the downstream direction by a flight distance significance greater than 15.
Only candidates with an impact parameter $\chi^2$, ${\rm IP} \chi^2$ 
(defined as the difference between the $\chi^2$ 
of the PV formed with and without the considered tracks) less than 25 are considered.
When more than one PV is reconstructed, that giving the smallest 
${\rm IP} \chi^2$ for the $B$ candidate is chosen. 
Tracks from selected candidates 
are required to have transverse momentum $p_{\rm T}$ satisfying $0.25<p_{\rm T}<40$\gevc
and \mbox{$p<$ 500\,GeV/$c$}.
Only \B candidates with decay times smaller than $9 \,\tau(\Bs)$~\cite{PDG2012} and 
with invariant mass in the range $[4900, 6000]\mevcc$ are kept.

Dimuon candidates from elastic diphoton production are heavily suppressed 
by requiring $p_{\rm T}(B) > 0.5 \gevc$.
The surviving background comprises mainly random combinations of muons from 
semileptonic decays of two different $b$ hadrons (\bbdim, where $X$ is any other set of particles).

Two channels, \BuJpsiK and $B^0 \to K^+ \pi^-$, serve as normalization modes. 
The first mode has trigger 
and muon identification efficiencies similar to those of the signal, but 
a different number of tracks in the final state. The second mode has a 
similar topology, but is triggered differently. The selection of 
these channels is as close as possible to that of 
the signal to reduce the impact of potential systematic uncertainties.

The \BdKpi selection is the same as for \Bmm signal except for muon identification.
The two tracks are nevertheless required to be
within the muon detector acceptance.

The $J/\psi \to \mu^+ \mu^-$ decay in the \BuJpsiK normalization 
channel is also selected similarly to the \Bmumu signals, except for the requirements
on the IP$\chi^2$ and mass. Kaon candidates are required to have ${\rm IP}\chi^2 >25$. 

A two-stage multivariate selection, based on boosted decision trees~\cite{Breiman,AdaBoost}
is applied to the \Bmm candidates.
A cut on the first multivariate discriminant, unchanged from Ref.~\cite{lhcbpaper3},
removes 80\,\% of the background while retaining 92\,\% of signal.
The efficiencies of this cut for the signal and the normalization samples are 
equal within 0.2\,\% as determined from simulation.

%--------------------------------
% Multivariate analysis
%--------------------------------
The output of the second multivariate discriminant, called BDT, 
and the dimuon invariant mass are used to classify
the selected candidates.
The nine variables entering the BDT are
the \B candidate IP, the minimum IP$\chi^2$ of the two muons with respect to any PV, the
sum of the degrees of isolation of the muons 
(the number of good two-track vertices a muon can make with other tracks 
in the event), the \B candidate decay time, $p_{\rm T}$, 
and isolation~\cite{cdf_iso}, the distance of closest approach between the two muons,
the minimum $p_{\rm T}$ of the muons,
and the cosine of the angle between the muon momentum in the dimuon 
rest frame and the vector perpendicular to both the \B candidate 
momentum and the beam axis. 

The BDT discriminant is trained using simulated samples consisting of
\Bsmumu for signal and \bbdim for background. The BDT response
is defined such that it is approximately uniformly distributed between zero 
and one for signal events and peaks at zero for the background. 
The BDT response is independent of the invariant mass for signal 
inside the search window. 
The probability for a \Bmumu event to have a given BDT value is 
obtained from data using \BdKpi, $\pi^+\pi^-$ and \BspiK, $K^+ K^-$ 
exclusive decays selected as the signal events and triggered
independently of the tracks from $B^0_{(s)}$ candidates.

The invariant mass lineshape of the signal events 
is described by a Crystal Ball function~\cite{crystalball}.
The peak values for the \Bs and \Bd mesons, \mBs and \mBd, 
are obtained from the \BsKK and \BdKpi, \Bdpipi samples.
The resolutions are determined by combining the results obtained 
with a power-law interpolation between the measured resolutions of charmonium and 
bottomonium resonances decaying into two muons with those obtained with a fit of the 
mass distributions of 
\BdKpi, \Bdpipi and \BsKK samples.
The results are  $ \sigma_{\Bs}  =  25.0 \pm 0.4 \mevcc$ and
$\sigma_{\Bd}  =  24.6 \pm 0.4 \mevcc$, respectively.
The transition point of the radiative tail is obtained from simulated \Bsmumu events smeared
to reproduce the mass resolution measured in data.

%------------------------
% Normalization
%------------------------
The \Bsmumu and \Bdmumu yields are translated 
into branching fractions using
\begin{eqnarray}
\BRof \Bmumu &=&  
\frac{{\cal B}_{\rm norm}  \,{\rm \epsilon_{\rm norm}}\,f_{\rm norm} }{ N_{\rm norm}\,{\rm \epsilon_{sig}} \,f_{d(s)} } \times
N_{\Bmumu}  \nonumber \\
& = & \alpha^{\rm norm}_{\Bmumu} \times N_{\Bmumu},
\label{eq:normalization}
\end{eqnarray}
where ${\cal B}_{\rm norm}$ represents the branching fraction, 
$N_{\rm norm}$ the number of signal events in the normalization 
channel obtained from a fit to the invariant mass distribution, 
%and $\alpha^{\rm norm}_{\Bmumu}$  the normalization factors. Finally, 
and $N_{\Bmumu}$ is the number of observed signal events.

The factors $f_{d(s)}$ and $f_{\rm norm}$ indicate the probabilities
that a $b$ quark fragments into a $B^0_{(s)}$ meson and into the hadron involved
in the given normalization mode, respectively. 
We assume $f_d=f_u$ and use $f_s/f_d = 0.256 \pm 0.020$ 
measured in $pp$ collision data at $\sqrt{s}=7$ TeV~\cite{LHCb-PAPER-2012-037}.
This value is in agreement within $1.5 \,\sigma$ 
with that found at $\sqrt{s}=8$ TeV by comparing the ratios of the yields of 
\BsJpsiPhi and \BuJpsiK decays.
The measured dependence of $f_s/f_d$ on $p_{\rm T}(B)$~\cite{LHCb-PAPER-2012-037} 
is found to be negligible for this analysis.

The efficiency ${\rm \epsilon_{sig(norm)}}$ for the signal (normalization channel) is
the product of the reconstruction efficiency of the final state particles 
including the geometric detector acceptance, 
the selection efficiency and the trigger efficiency. 
%--
The ratio of acceptance, reconstruction and selection efficiencies is computed using 
simulation. Potential differences between data and simulation 
are accounted for as systematic uncertainties. 
%The selection efficiencies are determined using simulation and cross-checked with data.
Reweighting techniques are used for all the distributions in the simulation that  do not match 
those from data. The trigger efficiency is evaluated with data-driven techniques~\cite{tistos}.
%--
The observed numbers of \BuJpsiK and \BdKpi candidates in the 2012 dataset are
$424\,200\pm 1500$ and $14\,600\pm 1100$, respectively.
The two normalization factors $\alpha^{\rm norm}_{\Bmumu}$ are in agreement within the uncertainties, 
and their weighted average, taking correlations into account,
gives
$\alpha_{\Bsmumu}= (2.52 \pm 0.23) \times 10^{-10}$ and
$\alpha_{\Bdmumu}= (6.45 \pm 0.30) \times 10^{-11}$. 

%-------------------
% Mass/BDT binning
%-------------------
In total, 24\,044 muon pairs with invariant mass between 4900 and 6000\mevcc 
pass the trigger and selection requirements. 
Given the measured normalization factors and assuming the SM branching fractions, 
the data sample is expected to contain about 14.1 $B^0_{s}\to \mu^+\mu^-$  and 
$1.7$ $B^0 \to \mu^+\mu^-$ decays. 

The BDT range is divided into eight bins with boundaries $[0.0,0.25,0.4,0.5,0.6,0.7,0.8,0.9,1.0]$. 
For the 2012 dataset, only one bin is considered in the BDT range
0.8--1.0 due to the lack of events in the mass sidebands for ${\rm BDT} > 0.9$.
The signal regions are defined by $m_{B^{0}_{(s)}} \pm 60$\mevcc.

%---------------------------
% Mass range and background
%---------------------------
The expected number of combinatorial background events is determined 
by interpolating from the invariant mass sideband regions 
defined as $[4900 \mevcc, m_{B^0} - 60 \mevcc]$ and $[m_{B^0_s}+60 \mevcc, 6000 \mevcc]$.
The low-mass sideband and the \Bd and \Bs 
signal regions are potentially polluted by exclusive backgrounds with or without misidentification
of the muon candidates. 

The first category includes $B^0 \to \pi^- \mu^+ \nu_{\mu}$, \Bhh,
$B^0_s \to K^- \mu^+ \nu_{\mu}$ and $\Lambda^0_b \to p \mu^- \overline{\nu}_{\mu}$ decays.
The $B^0 \to \pi^- \mu^+ \nu_{\mu}$ and \Bhh branching fractions are taken from Ref.~\cite{PDG2012}.
The theoretical estimates of the $\Lambda^0_b \to p \mu^- \overline{\nu}_{\mu}$ and $B^0_s \to K^- \mu^+ \nu_{\mu}$ 
branching fractions are taken from Refs.~\cite{datta} and \cite{BsKmunu}, respectively.
The mass and BDT distributions
of these modes are evaluated from simulated samples where the $K\to\mu$, $\pi \to \mu$ 
and $p \to \mu$ misidentification probabilities as a function of momentum and transverse
momentum are those determined from $D^{*+} \to D^0 \pi^+, D^0 \to K^- \pi^+$ and $\Lambda \to p \pi^- $ 
data samples. We use the $\Lambda^0_b$ fragmentation fraction $f_{\Lambda^0_b}$  measured by LHCb~\cite{Aaij:2011jp} 
and account for its $p_{\rm T}$ dependence.

The second category includes \mbox{$B^+_c \to J/\psi(\mu^+ \mu^-) \mu^+ \nu_{\mu}$}, 
$B^0_s \to \mu^+ \mu^- \gamma$ and $B^{0(+)} \to \pi^{0(+)} \mu^+ \mu^-$ 
decays, evaluated assuming  branching fraction 
values from Refs.~\cite{Abe:1998wi}, \cite{Nikitin1}
and \cite{Bpimumu}, respectively. Apart from \Bhh, all background modes are
normalized relative to the \BuJpsiK decay.
The  $B^0 \to \pi^- \mu^+ \nu_{\mu}$, \Bhh and 
$B^{0(+)} \to \pi^{0(+)} \mu^+ \mu^-$ decays are the 
dominant exclusive modes in the range ${\rm BDT} >0.8$, which accounts for 70\,\% 
of the sensitivity.

In the full BDT range, $ 8.6 \pm 0.7$ 
doubly misidentified \Bhh decays are expected in the full 
mass interval, $4.1^{+1.7}_{-0.8}$ in the \Bd and 
$0.76^{+0.26}_{-0.18}$ in the \Bs signal region. The expected yields for 
$\Bd \to \pi^- \mu^+ \nu_{\mu}$ 
and $\B^{0 (+)} \to \pi^{0(+)} \mu^+ \mu^-$ 
are $41.1 \pm 0.4$ and $11.9 \pm 3.5$, respectively, in the full mass and BDT ranges. 
The contributions 
of these two backgrounds above $m_{\Bd}-60 \mevcc$ are negligible.
The fractions of these backgrounds with ${\rm BDT}>0.8$, in the full mass range, are $(19.0 \pm 1.4)\,\%$, $(11.1 \pm 0.5)\,\%$, 
and $(12.2 \pm 0.3)\,\%$ for \Bhh, $\Bd \to \pi^- \mu^+ \nu_{\mu}$ and 
$\B^{0 (+)} \to \pi^{0(+)} \mu^+ \mu^-$ decays, respectively.

A simultaneous unbinned maximum-likelihood fit to the mass projections in the BDT bins is performed
on the mass sidebands to determine the number of expected combinatorial background events 
in the \Bd and \Bs signal regions used in the derivation of the branching fraction limit.
In this fit the parameters that describe the mass distributions of the exclusive backgrounds, 
their fractional yields in each BDT bin and their overall yields are limited by Gaussian constraints
according to their expected values and uncertainties.
The combinatorial background is parameterized with an exponential function 
with slope and normalization allowed to vary.
The systematic uncertainty on the estimated number of combinatorial 
background events in the signal regions is determined 
by fluctuating the number of events observed in the sidebands according to a Poisson distribution, 
and by varying the exponential slope according to its uncertainty.
The same fit is then performed on the full mass range 
to determine the \Bsmumu and \Bdmumu branching fractions, which are free parameters of the fit.
The \Bsmumu and \Bdmumu  fractional yields in BDT bins are constrained to the BDT fractions calibrated with the
\Bhh sample. The parameters of the Crystal Ball functions 
that describe the mass lineshapes and the normalization factors 
are restricted by Gaussian constraints according to their expected values and uncertainties.

%--------------
% Results
%--------------
The compatibility of the observed distribution of events  
with that expected for a given branching fraction 
hypothesis is computed using the \CLs method~\cite{Read_02}.
The method provides \CLsb, a measure of the 
compatibility of the observed distribution with the signal plus background 
hypothesis, \CLb, a measure of the compatibility with the background-only 
hypothesis, and \mbox{$\CLs=\CLsb/\CLb$}.

The invariant mass signal regions are divided into nine bins 
with boundaries $m_{B^{0}_{(s)}} \pm 18, 30, 36, 48, 60$\mevcc.
In each bin of the two-dimensional space formed by the dimuon mass and the 
BDT output we count the number of observed candidates, and compute the expected number 
of signal and background events.

The comparison of the distributions of observed events and expected
background events in the 2012 dataset results in p-values \mbox{$(1-\CLb)$} of $9 \times 10^{-4}$ 
for the \Bsmumu and 0.16 for the \Bdmumu decay, 
computed at the branching fraction values corresponding to $\CLsb=0.5$.
We observe an excess of \Bsmumu candidates with respect to background 
expectation with a significance of 3.3 standard deviations. 
The simultaneous unbinned maximum-likelihood fit gives 
$\BRof \Bsmumu = (5.1^{\,+2.3}_{\,-1.9}({\rm stat}) ^{\,+0.7}_{\,-0.4} ({\rm syst})) \times 10^{-9}$. 
The statistical uncertainty reflects the interval corresponding to a change of 0.5 with respect to the
maximum of the likelihood after fixing all the fit parameters to their expected values
except the \Bsmumu and \Bdmumu branching fractions and the slope and normalization 
of the combinatorial background.
The systematic uncertainty is obtained by subtracting in quadrature the statistical uncertainty 
from the total uncertainty obtained from the likelihood with all nuisance parameters left to vary
according to their uncertainties. An additional systematic uncertainty 
of $0.16 \times 10^{-9}$ reflects the impact on the result of the change in 
the parameterization of the combinatorial background from a single to a double exponential, 
and is added in quadrature.

%-- 2012 results:
The expected and measured limits on the \Bdmumu branching fraction at 90\,\% 
and 95\,\% \CL are shown in Table~\ref{tab:b0limit}.
The expected limits are computed allowing for the 
presence of \Bmumu events according to the SM branching fractions, including 
cross-feed between the two modes. 

\begin{table}[thb]
\caption{Expected and observed limits on the \Bdmumu branching fractions for the 2012  and
for the combined 2011+2012 datasets.}
\label{tab:b0limit}
\begin{center}
\newcommand\Sp{\rule{2.5mm}{0pt}}
\begin{tabular}{llrr}
\hline\hline 
  Dataset \TTstrut\BBstrut & Limit at & 90\,\% \CL & 95\,\% \CL\\ 
\hline 
 2012 \TTstrut      & Exp. bkg+SM     &  $8.5 \times 10^{-10}$  & $10.5 \times 10^{-10}$\\ 
                    & Exp. bkg        &  $7.6 \times 10^{-10}$  & $9.6 \times 10^{-10}$\\
                    & Observed        &  $10.5\times 10^{-10}$  & $12.5 \times 10^{-10}$\\ 
\hline \hline
2011+2012 \TTstrut   & Exp. bkg+SM    &  $5.8 \times 10^{-10}$ & $7.1 \times 10^{-10}$\\ 
                     & Exp. bkg       &  $5.0 \times 10^{-10}$ & $6.0 \times 10^{-10}$\\ 
                     & Observed       &  $8.0 \times 10^{-10}$ & $9.4 \times 10^{-10}$\\ 
\hline \hline
\end{tabular}
\end{center}
\end{table}

% -- analysis of the 2011 dataset:
The contribution of the exclusive background components 
is also evaluated for the 2011 dataset, modifying the number
of expected combinatorial background in
the signal regions. The results for the \Bmumu branching fractions have been updated accordingly.
We obtain  \BRof \Bsmumu $< 5.1 \times 10^{-9}$ and \BRof \Bdmumu $< 13 \times 10^{-10}$ at 95\,\% CL to 
be compared to the published limits \BRof \Bsmumu $< 4.5 \times 10^{-9}$ and \BRof \Bdmumu $<10.3 \times 10^{-10}$ 
at 95\,\% CL~\cite{lhcbpaper3}, respectively.
The (1-\CLb) p-value for \Bsmumu changes from 18\,\% to 11\,\% and the \Bsmumu branching fraction 
increases by $\sim 0.3 \,\sigma$ from $(0.8^{\,+1.8}_{\,-1.3}) \times 10^{-9}$ 
to  $(1.4^{\,+1.7}_{\,-1.3}) \times 10^{-9}$. 
This shift is compatible with the systematic uncertainty previously assigned to 
the background shape~\cite{lhcbpaper3}.
The values of the \Bsmumu branching fraction obtained with the 2011 and 2012 datasets are compatible 
within $1.5\,\sigma$.

% -- combination of 2011 and 2012 data:
The 2011 and 2012 results are combined by computing the \CLs 
and performing the maximum-likelihood fit simultaneously 
to the eight and seven BDT bins of the 2011 and 2012 datasets, respectively. 
The parameters that are considered 100\,\% correlated between the two datasets 
are $f_s/f_d$, \BRof \BuJpsiK and \BRof \BdKpi, 
the transition point of the Crystal Ball function describing the 
signal mass lineshape,
the mass distribution of the \Bhh background, the BDT and mass distributions 
of the \BdPiMuNu and \bpimumu backgrounds and the SM predictions 
of the \Bsmumu and \Bdmumu branching fractions.
The distribution of the expected and observed events in bins of BDT 
in the signal regions obtained from the simultaneous analysis 
of the 2011 and 2012 datasets, are summarized  in Table~\ref{tab:data_bsdmm_2011_2012}.

\begin{figure}[!tb]
\centering
\includegraphics[width=0.45\textwidth]{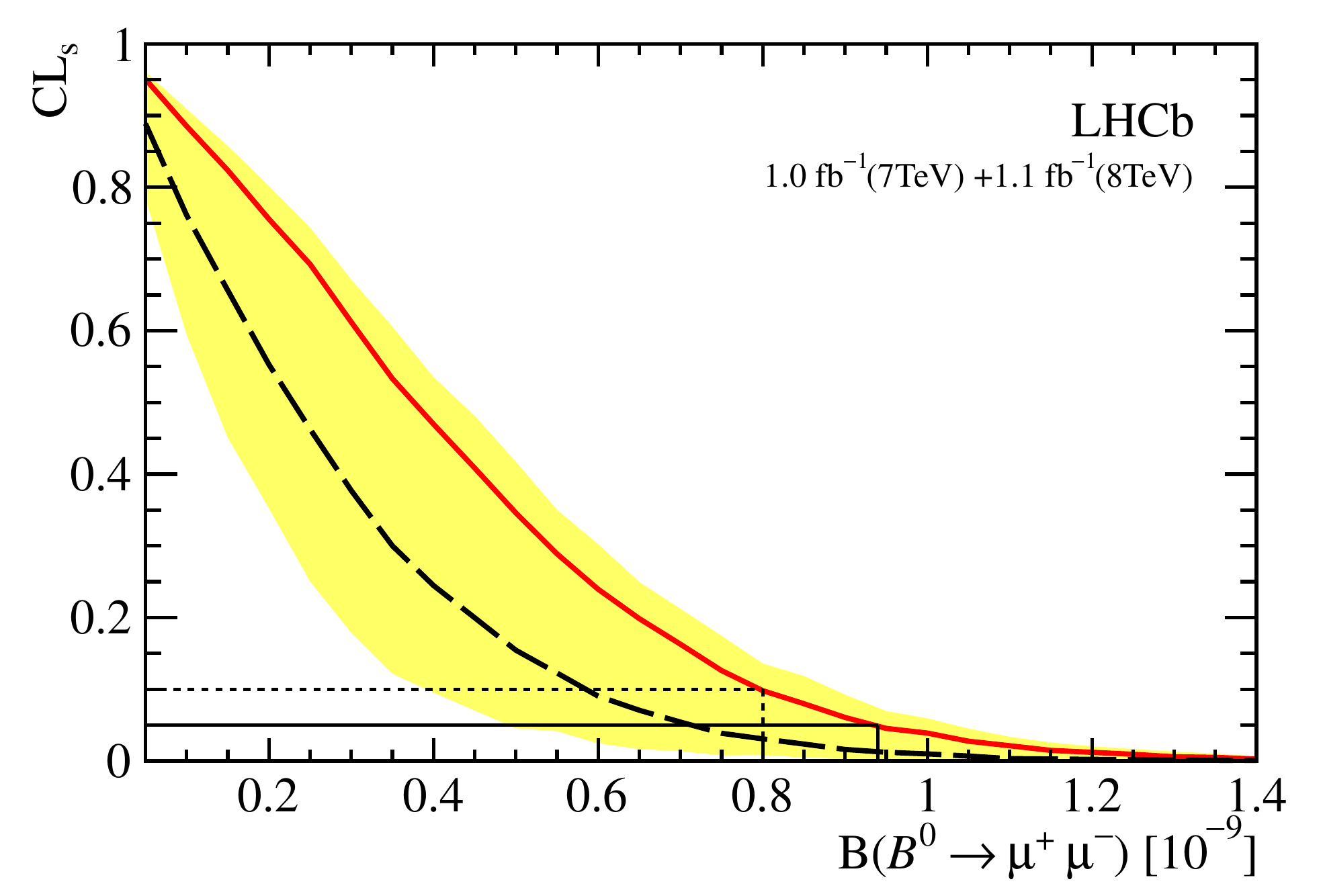}
\vspace{-4mm}
\caption
{ \CLs as a function of the assumed \Bdmumu branching fraction for the combined 2011+2012 dataset.
The dashed gray curve is the median of the expected \CLs\ distribution if 
background and SM signal were observed.
The shaded yellow area covers, for each branching fraction value, 34\,\% of the 
expected \CLs distribution on each side of its median.
The solid red curve is the observed \CLs.}
\label{fig:cls_bd}
\end{figure}

\begin{figure}[t]
  \begin{center}
    \includegraphics*[width=0.45\textwidth]{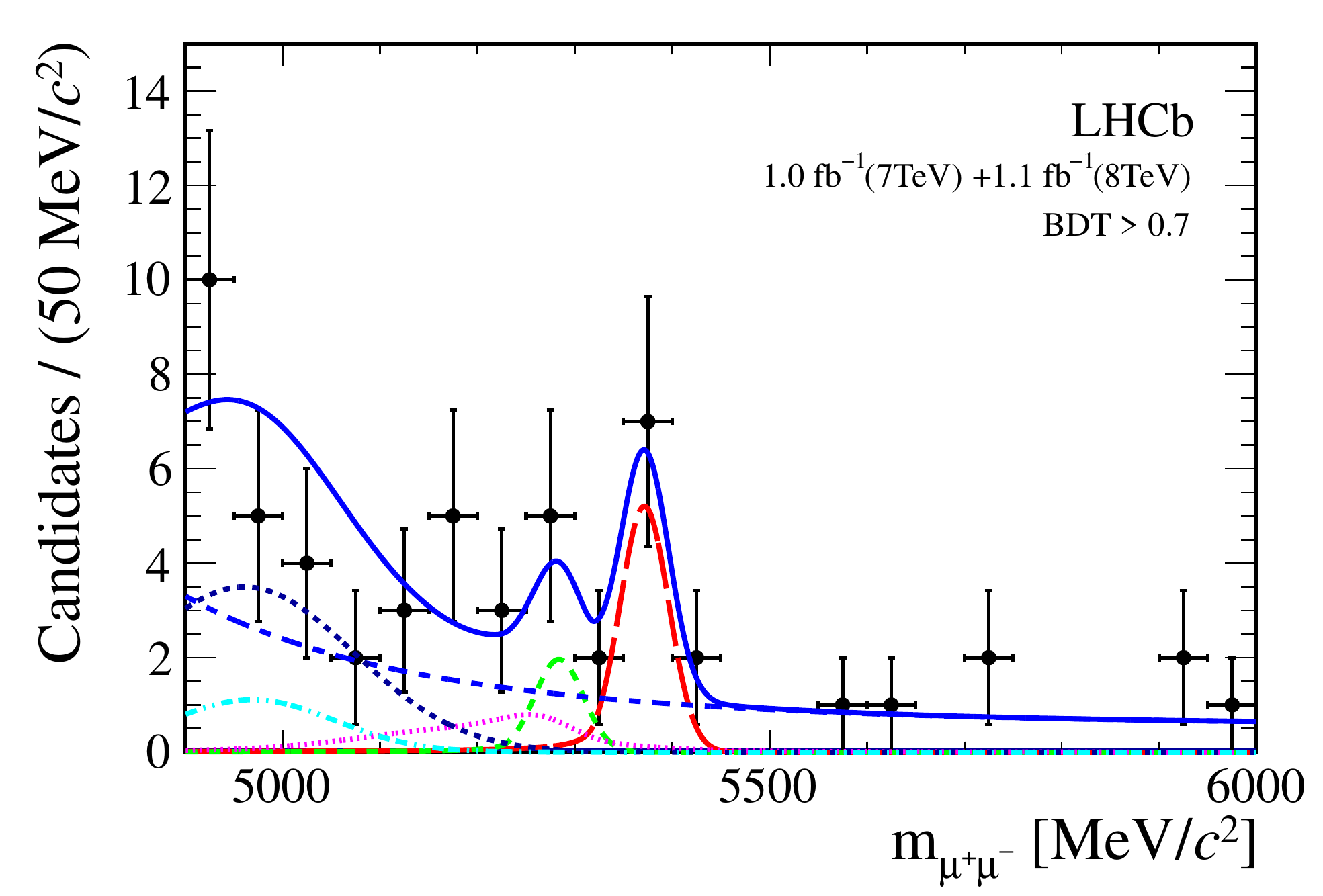}
  \end{center}
\caption{
Invariant mass distribution of the selected \Bsmumu candidates (black dots) with ${\rm BDT}>0.7$ 
in the combined 2011+2012 dataset.
The result of the fit is overlaid (blue solid line) and the different components detailed:
\Bsmumu  (red long dashed), \Bdmumu (green medium dashed), \Bhh (pink dotted), \BdPiMuNu (black short dashed) and \bpimumu 
(light blue dot dashed), and the combinatorial background (blue medium dashed).}
\label{fig:mass}
\end{figure}

The expected and observed upper limits for the \Bdmumu channel 
obtained from the combined 2011+2012 datasets are 
summarized in Table~\ref{tab:b0limit} and the expected and observed \CLs values
as a function of the branching fraction are shown in Fig.~\ref{fig:cls_bd}.
The observed \CLb value at \CLsb= 0.5 is 89\,\%.
%--
%--
The probability that background processes can produce the observed number of \Bsmumu candidates or more
is  $5 \times 10^{-4}$  
and corresponds to a statistical significance of $3.5 \,\sigma$.
The value of the \Bsmumu branching fraction obtained from the fit is
\[
\BRof \Bsmumu = (3.2^{\,+1.4}_{\,-1.2} ({\rm stat}) ^{\,+0.5}_{\,-0.3} ({\rm syst}) ) \times 10^{-9}
\]
and is in agreement with the SM expectation.
The invariant mass distribution of the \Bmumu candidates with
${\rm BDT}>0.7$ is shown in Fig.~\ref{fig:mass}. 
%
%---
\begin{center}
\begin{table*}[htb]
\centering
\caption[]{Expected combinatorial background,  \Bhh peaking background, cross-feed, and signal events assuming the SM prediction, together with the number of observed candidates in the \Bsmumu and \Bdmumu mass signal regions, in bins of BDT for the 2011 (top) and for the 2012 (bottom) data samples. The quoted errors include statistical and systematic uncertainties.} 
    \label{tab:data_bsdmm_2011_2012}
\begin{tabular}{llcccccccc}
\hline\hline
Mode & BDT bin & 0.0 -- 0.25 & 0.25 -- 0.4 & 0.4 -- 0.5 & 0.5 -- 0.6 & 0.6 -- 0.7 & 0.7 -- 0.8 & 0.8 -- 0.9 & 0.9 -- 1.0 \TTstrut\BBstrut \\
\hline
\Bsmumu 
 & Exp. comb. bkg & $1880^{+33}_{-33}$ & $55.5^{+3.0}_{-2.9}$ & $12.1^{+1.4}_{-1.3}$ & $4.16^{+0.88}_{-0.79}$ & $1.81^{+0.62}_{-0.51}$ & $0.77^{+0.52}_{-0.38}$ & $0.47^{+0.48}_{-0.36}$ & $0.24^{+0.44}_{-0.20}$  \TTstrut\\
(2011) & Exp. peak. bkg & $0.13^{+0.07}_{-0.05}$ & $0.07^{+0.02}_{-0.02}$ & $0.05^{+0.02}_{-0.02}$ & $0.05^{+0.02}_{-0.01}$ & $0.05^{+0.02}_{-0.01}$ & $0.05^{+0.02}_{-0.01}$ & $0.05^{+0.02}_{-0.01}$ & $0.05^{+0.02}_{-0.01}$  \TTstrut\\
 & Exp. signal & $2.70^{+0.81}_{-0.80}$ & $1.30^{+0.27}_{-0.23}$ & $1.03^{+0.20}_{-0.17}$ & $0.92^{+0.15}_{-0.13}$ & $1.06^{+0.17}_{-0.15}$ & $1.10^{+0.17}_{-0.15}$ & $1.26^{+0.20}_{-0.17}$ & $1.31^{+0.28}_{-0.25}$  \TTstrut\\
 &  Observed & $1818$ & $39$ & $12$ & $6$ & $1$ & $2$ & $1$ & $1$  \TTstrut\Bstrut\\
\hline
\Bdmumu 
        & Exp. comb. bkg & $1995^{+34}_{-34}$ & $59.2^{+3.3}_{-3.2}$ & $12.6^{+1.6}_{-1.5}$ & $4.44^{+0.99}_{-0.86}$ & $1.67^{+0.66}_{-0.54}$ & $0.75^{+0.58}_{-0.40}$ & $0.44^{+0.57}_{-0.38}$ & $0.22^{+0.48}_{-0.20}$  \TTstrut\\
 (2011) & Exp. peak. bkg & $0.78^{+0.38}_{-0.29}$ & $0.40^{+0.14}_{-0.10}$ & $0.31^{+0.11}_{-0.08}$ & $0.28^{+0.09}_{-0.07}$ & $0.31^{+0.10}_{-0.08}$ & $0.30^{+0.10}_{-0.07}$ & $0.31^{+0.10}_{-0.08}$ & $0.30^{+0.11}_{-0.08}$  \TTstrut\\
 & Exp. cross-feed & $0.43^{+0.13}_{-0.13}$ & $0.21^{+0.04}_{-0.04}$ & $0.16^{+0.03}_{-0.03}$ & $0.15^{+0.03}_{-0.02}$ & $0.17^{+0.03}_{-0.03}$ & $0.17^{+0.03}_{-0.02}$ & $0.20^{+0.03}_{-0.03}$ & $0.21^{+0.05}_{-0.04}$  \TTstrut\\
 & Exp. signal & $0.33^{+0.10}_{-0.10}$ & $0.16^{+0.03}_{-0.03}$ & $0.13^{+0.02}_{-0.02}$ & $0.11^{+0.02}_{-0.02}$ & $0.13^{+0.02}_{-0.02}$ & $0.13^{+0.02}_{-0.02}$ & $0.15^{+0.02}_{-0.02}$ & $0.16^{+0.03}_{-0.03}$  \TTstrut\\
  & Observed & $1904$ &  $50$ &  $20$ &  $5$ &  $2$ &  $1$ &  $4$ &  $1$  \TTstrut\Bstrut\\
\hline \hline
%-
Mode & BDT bin & 0.0 -- 0.25 & 0.25 -- 0.4 & 0.4 -- 0.5 & 0.5 -- 0.6 & 0.6 -- 0.7 & 0.7 -- 0.8 
& \multicolumn{2}{c}{0.8--1.0} \TTstrut\BBstrut \\
\hline
\Bsmumu
 & Exp. comb. bkg & $2345^{+40}_{-40}$ & $56.7^{+3.0}_{-2.9}$ & $13.1^{+1.5}_{-1.4}$ & $4.42^{+0.91}_{-0.81}$ & $2.10^{+0.67}_{-0.56}$ & $0.35^{+0.42}_{-0.22}$ & \multicolumn{2}{c}{$0.39^{+0.33}_{-0.21}$}   \TTstrut\\
 (2012) & Exp. peak. bkg & $0.250^{+0.08}_{-0.07}$ & $0.15^{+0.05}_{-0.04}$ & $0.08^{+0.03}_{-0.02}$ & $0.08^{+0.02}_{-0.02}$ & $0.07^{+0.02}_{-0.02}$ & $0.06^{+0.02}_{-0.02}$ & \multicolumn{2}{c}{$0.10^{+0.03}_{-0.03}$}   \TTstrut\\
 & Exp. signal & $3.69^{+0.59}_{-0.52}$ & $2.14^{+0.37}_{-0.33}$ & $1.20^{+0.21}_{-0.18}$ & $1.16^{+0.18}_{-0.16}$ & $1.17^{+0.18}_{-0.16}$ & $1.15^{+0.19}_{-0.17}$ & \multicolumn{2}{c}{$2.13^{+0.33}_{-0.29}$}   \TTstrut\\
 & Observed & $2274$ & $65$ & $19$ & $5$ & $3$ & $1$ & \multicolumn{2}{c}{$3$}  \TTstrut\Bstrut\\
\hline
\Bdmumu 
 & Exp. comb. bkg & $2491^{+42}_{-42}$ & $59.5^{+3.3}_{-3.2}$ & $13.9^{+1.6}_{-1.5}$ & $4.74^{+1.00}_{-0.89}$ & $2.10^{+0.74}_{-0.61}$ & $0.55^{+0.50}_{-0.31}$ & \multicolumn{2}{c}{$0.29^{+0.34}_{-0.19}$}   \TTstrut\\
 (2012) & Exp. peak. bkg & $1.49^{+0.50}_{-0.36}$ & $0.86^{+0.29}_{-0.22}$ & $0.48^{+0.16}_{-0.12}$ & $0.44^{+0.15}_{-0.11}$ & $0.42^{+0.14}_{-0.10}$ & $0.37^{+0.13}_{-0.09}$ & \multicolumn{2}{c}{$0.62^{+0.21}_{-0.15}$}   \TTstrut\\
 & Exp. cross-feed & $0.63^{+0.10}_{-0.09}$ & $0.36^{+0.07}_{-0.06}$ & $0.20^{+0.04}_{-0.03}$ & $0.20^{+0.03}_{-0.03}$ & $0.20^{+0.03}_{-0.03}$ & $0.20^{+0.03}_{-0.03}$ & \multicolumn{2}{c}{$0.36^{+0.06}_{-0.05}$}   \TTstrut\\
 & Exp. signal & $0.44^{+0.06}_{-0.06}$ & $0.26^{+0.04}_{-0.04}$ & $0.14^{+0.02}_{-0.02}$ & $0.14^{+0.02}_{-0.02}$ & $0.14^{+0.02}_{-0.02}$ & $0.14^{+0.02}_{-0.02}$ & \multicolumn{2}{c}{$0.26^{+0.04}_{-0.03}$}   \TTstrut\\
 & Observed & $2433$ & $59$ & $19$ & $3$ & $2$ & $2$ & \multicolumn{2}{c}{$2$}  \TTstrut\Bstrut\\
\hline \hline
\end{tabular}
\end{table*}
 \end{center}

The true value of the \Bsmumu branching fraction is contained in the interval
$[1.3,5.8]\times 10^{-9} ( [1.1, 6.4] \times 10^{-9})$ at 90\,\% \CL (95\,\% \CL), 
where the lower and upper limit
are the branching fractions evaluated at 
\CLsb= 0.95 (\CLsb= 0.975) and \CLsb= 0.05 (\CLsb= 0.025), respectively.
These results are in good agreement with the lower and upper limits 
derived from integrating the profile likelihood obtained from the unbinned fit.

In summary, a search for the rare decays \Bsmumu
and \Bdmumu is performed using 1.0\invfb and 1.1\invfb of $pp$ collision data collected 
at $\sqrt{s} = 7$ TeV and $\sqrt{s}=8$ TeV, respectively.
The data in the \Bd search window are consistent with the background expectation and the world's best 
upper limit of \BRof \Bdmumu $< 9.4 \times 10^{-10}$ at 95\,\% CL is obtained.
The data in the \Bs search window show an excess of events with respect to the background-only prediction
with a statistical significance of $3.5 \,\sigma$. 
A fit to the data leads to  \BRof \Bsmumu $= (3.2^{\,+1.5}_{\,-1.2}) \times 10^{-9}$ which is in agreement
with the SM prediction. This is the first evidence for the decay \Bsmumu.

\section{Acknowledgements}
\noindent We express our gratitude to our colleagues in the CERN
accelerator departments for the excellent performance of the LHC. We
thank the technical and administrative staff at the LHCb
institutes. We acknowledge support from CERN and from the national
agencies: CAPES, CNPq, FAPERJ and FINEP (Brazil); NSFC (China);
CNRS/IN2P3 and Region Auvergne (France); BMBF, DFG, HGF and MPG
(Germany); SFI (Ireland); INFN (Italy); FOM and NWO (The Netherlands);
SCSR (Poland); ANCS/IFA (Romania); MinES, Rosatom, RFBR and NRC
``Kurchatov Institute'' (Russia); MinECo, XuntaGal and GENCAT (Spain);
SNSF and SER (Switzerland); NAS Ukraine (Ukraine); STFC (United
Kingdom); NSF (USA). We also acknowledge the support received from the
ERC under FP7. The Tier1 computing centres are supported by IN2P3
(France), KIT and BMBF (Germany), INFN (Italy), NWO and SURF (The
Netherlands), PIC (Spain), GridPP (United Kingdom). We are thankful
for the computing resources put at our disposal by Yandex LLC
(Russia), as well as to the communities behind the multiple open
source software packages that we depend on.

\bibliographystyle{LHCb}
\bibliography{bsmumureferences_2012}

\end{document}